\documentclass[epj]{webofc}
\usepackage[utf8]{inputenc}
\usepackage[varg]{txfonts}   
\usepackage{booktabs}
\usepackage{xcolor}
\definecolor{darkred}{rgb}{0.4,0.0,0.0}
\definecolor{darkgreen}{rgb}{0.0,0.4,0.0}
\definecolor{darkblue}{rgb}{0.0,0.0,0.4}
\usepackage[bookmarks,linktocpage,colorlinks,
    linkcolor = darkred,
    urlcolor  = darkblue,
    citecolor = darkgreen]{hyperref}
\usepackage{subfigure}
\wocname{EPJ Web of Conferences}
\woctitle{Lattice2017}
\begin{document}
%
\selectlanguage{english}
\title{%
Heavy light tetraquarks from Lattice QCD
}
\author{%
\firstname{Parikshit} \lastname{Junnarkar}\inst{1}\fnsep\thanks{Speaker, \email{parikshit@theory.tifr.res.in}} \and
\firstname{M} \lastname{Padmanath}\inst{2} \and
\firstname{Nilmani}  \lastname{Mathur}\inst{1}
}
\institute{%
Tata Institute of Fundamental Research, Mumbai 400005
\and
Instit\"ut  f\"ur  Theoretische Physik, Universit\"at Regensburg, D-93040 Regensburg, Germany.
}
\abstract{%
 We present preliminary results from a lattice calculation of  tetraquark states in the charm and bottom sector of the type $ud\bar{b}\bar{b}$, $us\bar{b}\bar{b}$, $ud\bar{c}\bar{c}$ and $sc\bar{b}\bar{b}$.  
 These calculations are performed on $N_f = 2 + 1 + 1$ MILC ensembles with lattice spacing of  $a = 0.12~\mathrm{fm} $ and $a=0.06~\mathrm{fm} $.
 A relativistic action with overlap fermions is employed for the light and charm quarks while a non-relativistic action with non-perturbatively improved coefficients is used in the bottom sector.
 Preliminary results provide a clear indication of presence of energy levels below the relevant thresholds of different tetraquark states.
 While in double charm sector we find shallow bound levels, our results suggest deeply bound levels with double bottom tetraquarks.
}
\maketitle
\section{Introduction}\label{intro}
The discovery of the resonances $Z_b(10607)$ \& $Z_b(10650)$ by BELLE \cite{Belle:2011aa} in 2012 has shown the existence of multiquark exotic states in the bottom sector.
Eventually the existence of a tetraquark state $Z_c(4430)$ was firmly established by the LHCb collaboration~\cite{Aaij:2014jqa}.
These new discoveries on the existence of a new bound state of matter have generated a lot of interest in exploring its hadronic structure with the leading candidate being that of a tetraquark state.
A tetraquark state, first employed by Jaffe \cite{Jaffe:1976ig} in the context of light scalar mesons and later for exotic spectroscopy \cite{Jaffe20051}, is a colour neutral state formed as a bound system of diquarks and antidiquarks. 
The tetraquark structure has been recently employed to identify favourable flavour, spin channels in the bottom sector. 
However, the tools employed in such searches are typically sum rule type calculations.
A first principles approach of lattice QCD is more desirable for such searches and in the past year, two lattice studies ~\cite{Bicudo:2016ooe,Francis:2016hui,Bicudo:2017szl} have identified a promising channel with the flavor content $ud\bar{b}\bar{b}$.
The calculation in Ref \cite{Bicudo:2016ooe} computed a potential of two heavy static antiquarks in presence of two light quarks using lattice QCD. 
This was then used to solve a coupled non-relativistic Schr{\"o}dinger equation to find a binding energy of $\Delta E = 90^{+43}_{-36}$ MeV in the $ud\bar{b}\bar{b}$ channel with $I(J^P) = 0(1^+)$.
In a later extension to this work in  Ref \cite{Bicudo:2017szl}, a resonance prediction was made by searching for the poles in the S and T matrices decaying in two B mesons.
The work in Ref \cite{Francis:2016hui} also confirmed a presence of a deeply bound level in the flavour channels of $ud\bar{b}\bar{b}$ and $us\bar{b} \bar{b}$ with the binding of 189(10) MeV and 98(7) MeV respectively.
In the work presented at the conference, we explore the tetraquarks of type $ud\bar{b}\bar{b}$, as in Ref \cite{Francis:2016hui}, confirming a presence of a level well below the threshold state and explore other flavour channels. 

\section{Operator setup}\label{ops}
We consider two types of operators here, namely a tetraquark operator with two quarks and two antiquarks having the desired quantum numbers and a two meson operator corresponding to the quantum numbers of that of the tetraquark state. 
The construction of the tetraquark operator employs a product of a diquark and antidiquark as suggested by Jaffe \cite{Jaffe20051}. 
The diquarks and antidiquarks are constructed with the so called ``good diquark" \cite{Jaffe20051} prescription.
We would like to construct a tetraquark operator with $I(J^P) = 0(1^+)$ in the diquark-antidiquark picture with two antibottom and two light quarks in the following configuration:
\begin{eqnarray}
&& (uq) \rightarrow (\overline{\mathbf{3}}_c,0,F_A), \quad \quad (\bar{b}\bar{b}) \rightarrow (\mathbf{3}_c,1,F_s), \nonumber \\
&& D(x) = u^a_{\alpha}(x) \  (C \gamma_5)_{\alpha \beta} \ q^{b}_{\beta}(x) \ \bar{b}^a_{\kappa}(x) (C \gamma_i)_{\kappa \rho} \ \bar{b}^b_{\rho}(x).
\end{eqnarray}
where the braces indicate the Color, Spin and Flavor (C,S,F) degrees of freedom.
For the case of light quarks $F_A$ indicates antisymmetric flavour combination which in this case will be in $\mathbf{3}_f$. For the double antibottom quarks, the flavour symmetry is manifestly symmetric $F_S$.
In the light diquark, the flavour $q \in (d,s,c)$ allows for studying different flavours of tetraquark states. 
The tetraquark operator shown on line two, indicated by $D(x)$ (keeping consistent notation with Ref \cite{Francis:2016hui}), is constructed by taking a dot product of the aforementioned diquark and antidiquark in color space.
A two meson operator with the quantum numbers as $I(J^P) = 0(1^+)$ can be constructed with different flavours $q \in (d,s,c) $ as :
\begin{eqnarray}
M^{d}(x) &=& B^+(x)B^{0*}(x) -B^0(x)B^{+*}(x) \rightarrow (q=d)    \nonumber \\
M^{s}(x) &=& B^+(x)B^{*}_s(x) -B^0_s(x)B^{+*}(x) \rightarrow (q=s) \nonumber \\
M^{c}(x) &=& B^+(x)B^{*}_c(x) -B^+_c(x)B^{+*}(x) \rightarrow (q=c) \nonumber \\
\end{eqnarray}
With these operators, correlation functions were computed on the lattices which will be described in the next section. 


\section{Lattice setup}\label{sec:lattice}
The calculations presented at the conference were performed on MILC ensembles employing HISQ gauge action and $N_f = 2 + 1 + 1$ flavours.
The ensemble parameters are shown in Table \ref{tab:pars}.
For both the ensembles
the charm and strange quark masses are tuned to their physical values, while 
the ratio $m_s/m_l$ is fixed to 5. 
The details can be found in Ref \cite{Bazavov:2012xda,Basak:2013oya}.
\begin{table}[htb]
  \centering
  \caption{\label{tab:pars}Ensemble parameters used in this work}
  \begin{tabular}{cccccc}\toprule
  V & $\beta$ & a(fm) & $m_\pi$(MeV) & $m_{\pi}L$ & $N_{\mathrm{confs}}$  \\\midrule
  $24^3 \times 64$ & 6.00 & 0.1207(14) & 305  & 4.54& 236 \\\midrule
  $48^3 \times 144$ & 6.72 & 0.0582(5) & 319 & 4.51 & 70 \\\bottomrule
  \end{tabular}
\end{table}

For the propagator computation, we have used wall sources \footnote{Propagators were also computed on point sources although the results are not shown here.} and the configurations were gauge fixed in Coulomb gauge and the links were then HYP smeared.
In the valence sector, we employ an overlap action where the details can be found in Refs \cite{Basak:2014kma,Mathur:2016hsm,Basak:2012py}. The use of overlap action eliminates $\mathcal{O}(a)$ lattice artifacts. 
In addition with the use of a multimass algorithm, a range of input bare masses can be accommodated.
The strange  quark mass is tuned by equating the fictitious pseudoscalar $\bar ss$ to 685 MeV \cite{Basak:2012py}.
The charm quark mass was tuned by setting its spin averaged kinetic mass $(m_{\eta_c} + 3 m_{J/\psi})/4$ to its physical value \cite{Basak:2014kma} and the bare values of the $am_c = 0.529, 0.290$ were used for the $24^3\times 64$ and $48^3 \times 144$ lattices respectively.

The bottom sector employs a NRQCD action as shown in Ref \cite{Lepage:1992tx}. 
In this set up, all terms up to $1/M^2_0$ and leading term in $1/M^3_0$ are included in the action where $M_0=am_b$ corresponds to the bare bottom quark mass.
The interaction part of NRQCD Hamiltonian includes $\mathcal{O}(a)$ improved derivatives and also six improvement coefficients $c_1..c_6$. 
The details of the action can be found in Ref \cite{Mathur:2016hsm}.
We use the non-perturbative determination of these coefficients as done by the HPQCD collaboration \cite{Dowdall:2011wh} for the coarser lattices. For the finer lattice,  they are set to their tree level values. 
The bottom quark mass was tuned by setting its lattice spin averaged mass of (1S) bottomonium: 
\begin{equation}
\overline{M}_{\mathrm{kin}}(1S) = \frac{3}{4} aM_{\mathrm{kin}}(\Upsilon) +  \frac{1}{4} aM_{\mathrm{kin}}(\eta_b), \quad \quad M_{\mathrm{kin}} = \frac{a^2p^2-(a\Delta E)^2}{2a\Delta E}
\end{equation}
to its experimental value. The kinetic mass is computed as shown in the right equation above.


\section{Results}\label{sec:results}

In obtaining the ground states of the correlation functions, we employ the variational method \cite{Luscher:1990ck,Blossier:2009kd}. 
With the tetraquark operator $D(x)$ and the two meson operator $M(x)$, we compute a correlator matrix:
\begin{equation}
C_{ij}(t) = \sum_x  \langle 0 | \mathcal{O}_i(x,t) \mathcal{O}_j^\dagger(0,0) | 0 \rangle \quad \quad \mathcal{O}_i(x,t) \in \big\{D(x,t),M(x,t)\big\}
\end{equation}
The correlator matrix is a $2 \times 2$ matrix and we solve a generalised eigenvalue problem to obtain the principle (ground state) correlation function as:
\begin{equation}
C_{ij}(t+\Delta t) v_j(t) = \lambda (t) C_{ij}(t) v_j(t) \quad \quad m_{\mathrm{eff}}(t) = - \frac{\mathrm{log} \lambda (t)}{\Delta t} 
\end{equation}
where $\lambda(t),v_j(t)$ are the eigenvalue and eigenvectors from the solution of the GEVP. It is convenient to construct effective masses as shown in right equation above.
\subsection{Results for $ud\bar{b}\bar{b}$}
\begin{figure}[h]
   \hspace{-2cm}
  \includegraphics[scale=0.38]{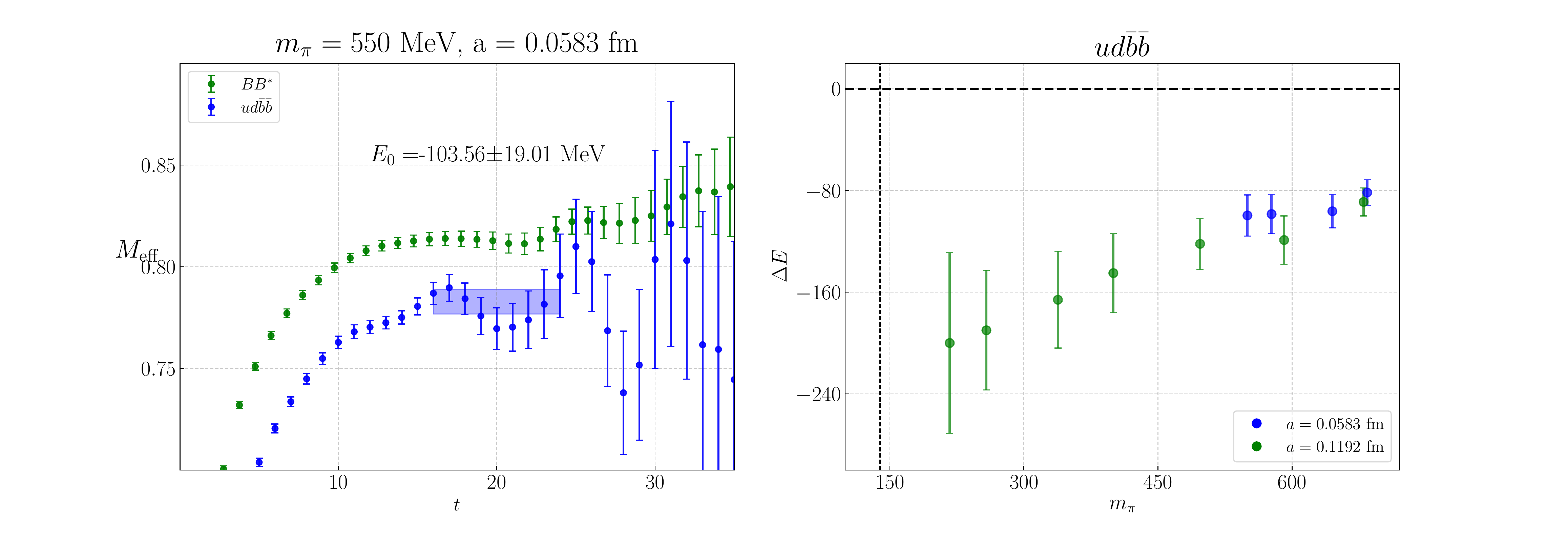}
  \caption{Left panel: Preliminary results for $ud\bar{b}\bar{b}$ tetraquark state. See text for the description of effective masses. Right panel : Binding energies for $a=0.0583$ fm (data in blue) and $a=0.1207$ fm (data in green) }
  \label{udbb}
\end{figure}
The results for the $ud\bar{b}\bar{b}$ tetraquarks are shown in Fig \ref{udbb}. 
In the case of the $ud\bar{b}\bar{b}$, the threshold states are those of two free mesons namely $B$ and $B^*$.
The plot in the left panel shows the effective masses of the two free meson states obtained from the correlator $C = C_B \* C_{B^*}$ (shown in green). 
The data in blue corresponds to the lowest level of the GEVP solution of the $2 \times 2 $ correlator matrix constructed from the correlation functions of the operators mentioned in section \ref{ops}.
The solution of the GEVP yields two levels, the excited level is found to be noisy and is not shown here.
With the use of wall sources, the ground states are seen to approach a plateau from below. 
The data in blue provides a clear indication of a level below the effective mass of the threshold correlator with the binding indicated on the plot. The results are also seen to be noisy for $t>25$. The calculation has been extended to smaller pion masses and the preliminary results are shown in the right panel. 
At the pion mass close to the SU(3) point, a comparison can be made with the results at the finer lattice spacing (right panel, data in blue) where the binding energies results are seen to be consistent.
As the pion mass is lowered, the binding is seen to get deeper, albeit with higher uncertainties.  This trend is found to be consistent with observations made in Ref \cite{Francis:2016hui}.
\subsection{Results for $us\bar{b}\bar{b}$}
\begin{figure}[h]
   \hspace{-2cm}
  \includegraphics[scale=0.38]{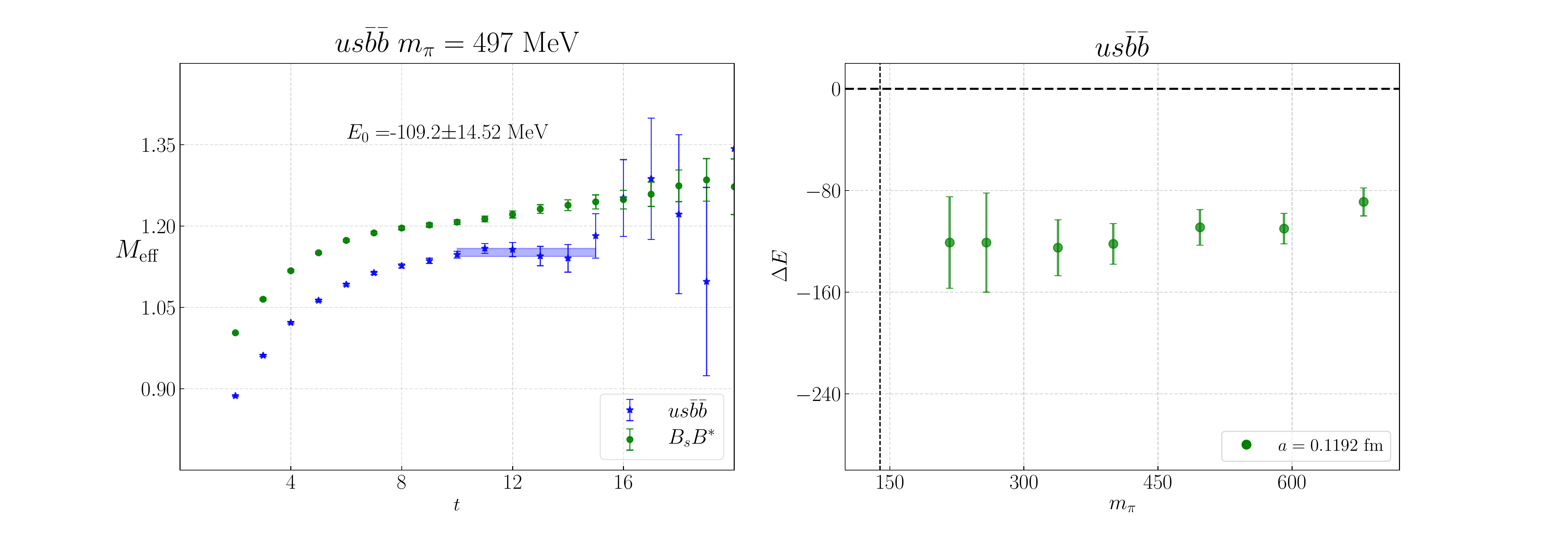}
  \caption{Preliminary results for $us\bar{b}\bar{b}$. Left panel : Effective energies for the threshold state and the lowest level of the GEVP solution.  Right panel : Summary of binding energies at lower pion masses at $a=0.1207$ fm.} \label{usbb}
\end{figure}
The results for $us \bar{b} \bar{b}$ tetraquarks are shown in Fig \ref{usbb}. 
The threshold  here is that of $B_s$ meson and $B^*$ meson. As before, the left panel indicates effective masses of the product of the correlators of $B_s$ and $B^*$ (shown in green) and the lowest level of the GEVP solution (data in blue). 
A clear indication of a level below the threshold state in seen for $m_\pi = 497$ MeV with the binding indicated on the plot.
The results shown here are computed on $24^3 \times 64$ lattice with $a=0.1207$ fm. 
The right panel in fig \ref{usbb} shows the results of the  pion mass dependence of the  binding energies where the slope of the binding energies with respect to pion masses is not as pronounced in comparison with the of $ud\bar{b} \bar{b}$ which possibly indicates a shallower binding at the physical point.
These results however are preliminary and will be improved upon in a future publication.
\subsection{Results for $sc\bar{b}\bar{b}$}
\begin{figure}[h]
   \centering
  \includegraphics[scale=0.4]{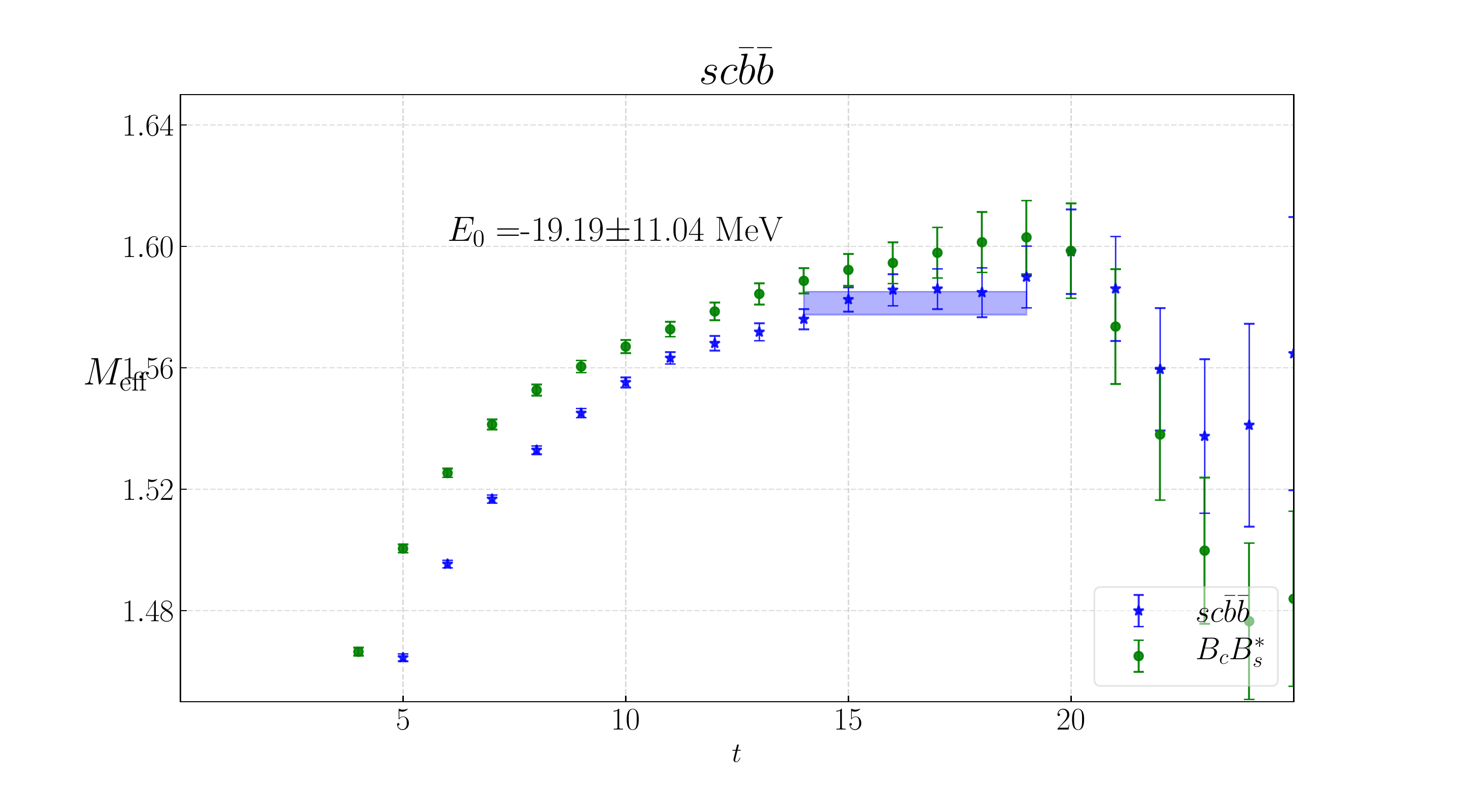}
  \caption{Preliminary results for $sc\bar{b}\bar{b}$ with all flavours at physical quark mass.}
  \label{scbb}
\end{figure}
We have also included the tetraquark state $sc\bar{b}\bar{b}$ in our calculation and the results shown in the Fig \ref{scbb}. As before the data in green indicates the effective mass of the threshold correlator which in this case is two free mesons namely $B_c$ meson and $B^*_s$ meson. 
We also note that this calculation was done with quark masses for all flavours at their physical value.  The main systematic in case will be the lattice spacing dependence of the binding energy which is currently ongoing. Due to the shallow result of the binding energy, study of finite volume effects in this case may also be important. 
\subsection{ Results for $ud\bar{c}\bar{c}$}
\begin{figure}[h]
  \includegraphics[scale=0.38]{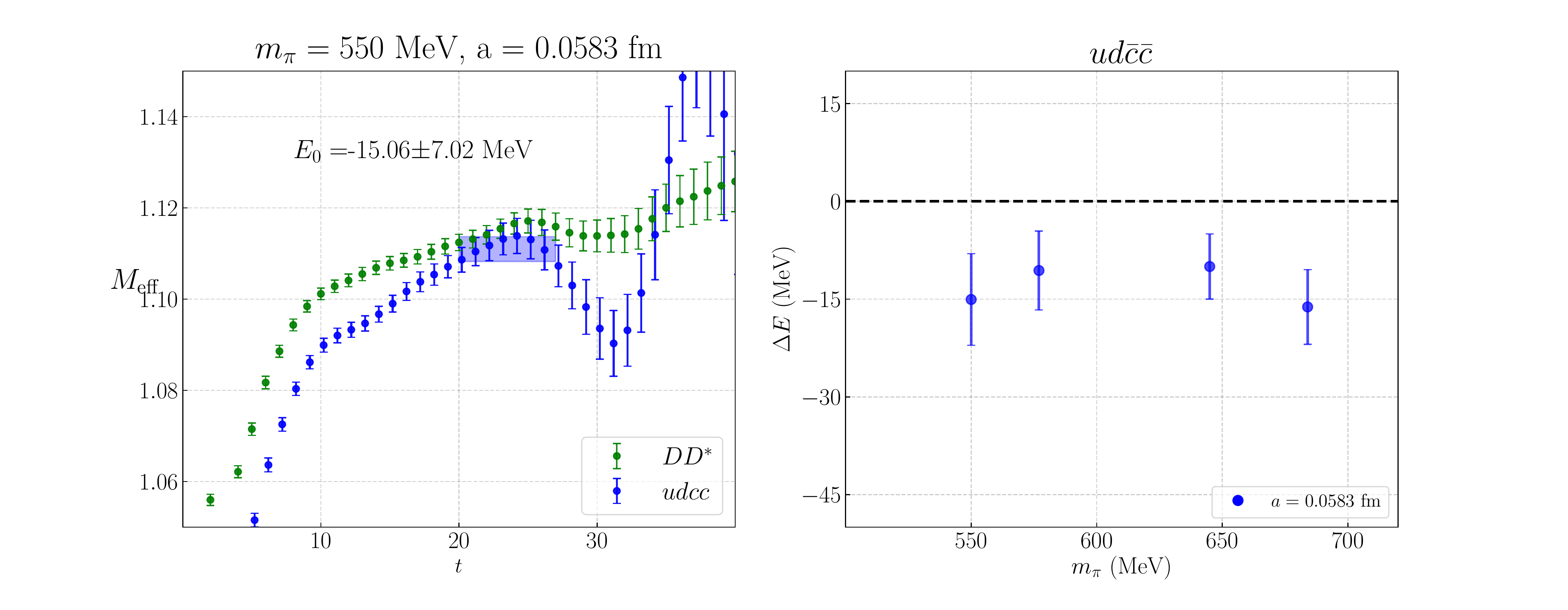}
  \caption{Preliminary results for $ud\bar{c}\bar{c}$. The color notation is the same as in previous plots.}
  \label{udcc}
\end{figure}
The charm analogue of the doubly bottom tetraquark state is the $ud\bar{c}\bar{c}$ state. The two meson thresholds in this case are the $D$ and $D^*$ mesons. The results of this calculation are shown in Fig \ref{udcc}. As before the data in green are the effective masses of the threshold correlator and the data in blue are those of the $ud\bar{c}\bar{c}$.
The results in this case are seen lie below but close to  the threshold of $D D^*$\footnote{ An error was later found in computation of $D^*$ mass. The corrected threshold is shown in Fig.~\ref{udcc}. } .
The results presented here are at lattice spacing $a = 0.0583$ fm and at heavier pion masses.
The extension to lower pion masses and another lattice spacing in currently underway. 
\section{Conclusions and outlook}
In this work, we have explored heavy light tetraquarks in the bottom and charm sector. The results on $a=0.1207$ fm is the progress since the conference.
In the $ud\bar{b}\bar{b}$ sector, we find a very clear indication of deeply bound levels and the binding energy increases as the pion mass is lowered.
The results shown here are preliminary and with added statistics these may change.
The results at lattice spacings $a=0.0583$ fm and $a=0.1207$ fm are seen to be consistent indicating no significant lattice spacing dependence.
The work is being currently extended to improve statistics and include results at lower pion masses for at least one more lattice spacing.
The results on $us\bar{b}\bar{b}$ are also seen to provide a clear indication of levels below the threshold state at various pion masses. 
The trend in the slope of the binding energy  approaching to the physical point is not as pronounced as that of $ud\bar{b}\bar{b}$ indicating that the binding may be shallower. 
This has also been noted in Ref \cite{Francis:2016hui}.
The effective mass of  $sc\bar{b}\bar{b}$ state is seen to be closer to the threshold at the physical values of the quark masses. 
The results on $ud\bar{c}\bar{c}$ are seen to lie below but close to the threshold indicating a shallow bound level in this channel.

\section{Acknowledgements}
The calculations are performed using computing resources of the Indian Lattice
Gauge Theory Initiative and the Department of Theoretical Physics, TIFR. 
We thank A. Salve, K. Ghadiali and P. Kulkarni for technical supports. 
P.M.J. and N.M. acknowledge support from the Department of Theoretical physics, TIFR. M. P. acknowledges support from Deutsche Forschungsgemeinschaft Grant No. SFB/TRR 55 and EU under grant no. MSCA-IF-EF-ST-744659 (XQCDBaryons).
We are grateful to the MILC collaboration and in particular 
to S. Gottlieb for providing us with the HISQ lattices. 

\bibliographystyle{JHEP-2}
\bibliography{tetraquarks_lattice2017}

\end{document}